\documentclass[twocolumn,prl,superscriptaddress,showpacs]{revtex4}
\usepackage{graphicx}
\usepackage{amsmath}
\newcommand{\figref}[1]{Fig.~\ref{#1}}
\begin{document}
%-------------------------------------------------------------------------
\title{Dressed Collective Qubit States and the Tavis-Cummings Model in Circuit QED}
\author{J.~M.~Fink}
\author{R.~Bianchetti}
\author{M.~Baur}
\author{M.~G\"{o}ppl}
\author{L. Steffen}
\author{S.~Filipp}
\author{P.~J.~Leek}
\affiliation{Department of Physics, ETH Zurich, CH-8093, Zurich,
Switzerland.}
\author{A.~Blais}
\affiliation{D\'epartement de Physique, Universit\'e de Sherbrooke,
Sherbrooke, Qu\'ebec J1K 2R1, Canada.}
\author{A.~Wallraff}
\affiliation{Department of Physics, ETH Zurich, CH-8093, Zurich,
Switzerland.}
%-------------------------------------------------------------------------
\date{\today}
%-------------------------------------------------------------------------
\begin{abstract}
We present an ideal
realization of the Tavis-Cummings model in the absence of atom number
and coupling fluctuations by embedding a discrete number of fully
controllable superconducting qubits at fixed positions into a
transmission line resonator.
Measuring the vacuum Rabi mode splitting with one, two and three
qubits strongly coupled to the cavity field, we explore both bright
and dark dressed collective multi-qubit states and observe the discrete
$\sqrt{N}$ scaling of the collective dipole coupling strength. Our experiments
demonstrate a novel approach to explore collective states, such as
the $W$-state, in a fully globally and locally
controllable quantum system. Our scalable approach is interesting
for solid-state quantum information processing and for
fundamental multi-atom quantum optics experiments with fixed atom numbers.\end{abstract}
%-------------------------------------------------------------------------
\maketitle
%-------------------------------------------------------------------------
In the early 1950's, Dicke realized that under certain conditions a
gas of radiating molecules shows the collective behavior of a single
quantum system \cite{Dicke1954}. The idealized situation in which
$N$ two-level systems with identical dipole coupling are resonantly
interacting with a single mode of the electromagnetic field was
analyzed by Tavis and Cummings \cite{Tavis1968}.  This model
predicts the collective $N$-atom interaction strength to be
$G_\textrm{N}=g_\textrm{j}\sqrt{N}$, where $g_\textrm{j}$ is the
dipole coupling strength of each individual atom $j$. In fact, in
first cavity QED experiments the normal mode splitting, observable
in the cavity transmission spectrum \cite{Agarwal1984,Leslie2004},
was demonstrated with on average $\bar{N}>1$ atoms in optical
\cite{Raizen1989,Zhu1990} and microwave \cite{Bernardot1992}
cavities to overcome the relatively weak dipole coupling
$g_\textrm{j}$. The $\sqrt{N}$ scaling has been observed in the
regime of a small mean number of atoms $\bar{N}$ with dilute atomic
beams \cite{Bernardot1992,Childs1996,Thompson1998} and fountains
\cite{Munstermann2000} crossing a high-finesse cavity. In these
experiments, spatial variations of the atom positions and Poissonian
fluctuations in the atom number inherent to an atomic beam
\cite{Childs1996,Carmichael1999,Leslie2004} are unavoidable. In a
different limit where the cavity was populated with a very large
number of ultra-cold $^{87}\textrm{Rb}$ atoms \cite{Tuchman2006} and
more recently with Bose-Einstein condensates
\cite{Brennecke2007,Colombe2007} the $\sqrt{N}$ nonlinearity was
also demonstrated. However, the number of interacting atoms is
typically only known to about $\sim10$\% \cite{Brennecke2007}.

\begin{figure}[b]
\includegraphics[width=0.95 \columnwidth]{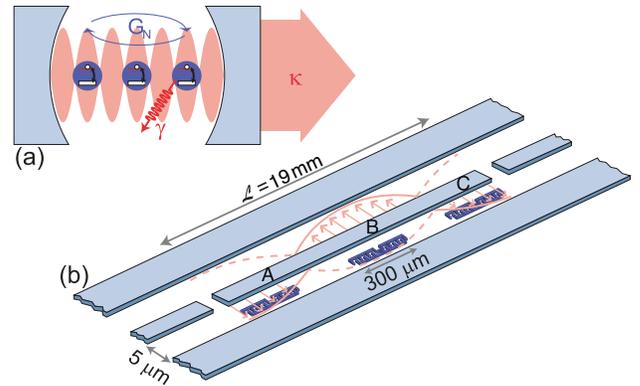}
\footnotesize \caption{Schematic of the experimental
set-up. (a) Optical analog. Three two-state atoms are identically
coupled to a cavity mode with photon decay rate $\kappa$, atomic energy
relaxation rate $\gamma$ and collective coupling strength
$G_\textrm{N}$. (b) Schematic of the investigated system. The
coplanar waveguide resonator is shown in light blue, the transmon
qubits A, B and C in violet and the first harmonic of the standing
wave electric field in red.} \label{setup}
\end{figure}

\begin{figure}[b]
\includegraphics[width=0.75 \columnwidth]{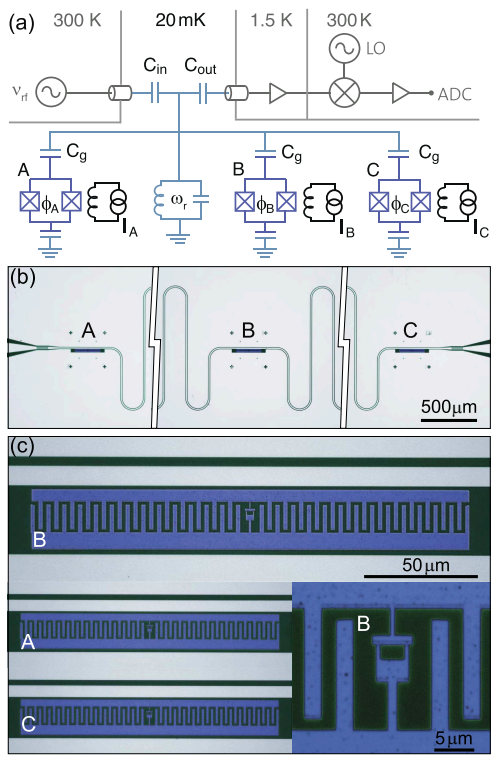}
\caption{Circuit diagram and false color optical images of
the sample. (a) Simplified electrical circuit diagram of the
experimental setup. The waveguide resonator operated at a
temperature of 20 \textrm{mK}, indicated as LC oscillator with
frequency $\omega_r$, is coupled to input and output leads with the
capacitors $C_\textrm{in}$ and $C_\textrm{out}$. Qubits A, B and C
are controlled with external current biased coils ($I_{A,B,C}$) and
coupled to the resonator via identical capacitors $C_\textrm{g}$. A
transmission measurement is performed by applying a measurement tone
$\nu_\textrm{rf}$ to the input port of the resonator, amplifying the
transmitted signal and digitizing it with an analog-to-digital
converter (ADC) after down-conversion with a local oscillator (LO)
in a heterodyne detection scheme. (b) The coplanar microwave
resonator is shown truncated in gray (substrate in dark green) and
the locations of qubits A, B and C are indicated. (c) Top, magnified
view of transmon qubit B (violet) embedded between ground plane and
center conductor of the resonator. Bottom left, qubits A and C, of same dimensions as qubit B, are shown at reduced scale. Bottom right, magnified view of SQUID loop of qubit B.}
\label{sample}
\end{figure}

Here we present an experiment in which the Tavis-Cummings model is
studied for a discrete set of fully controllable artificial atoms at fixed positions
and with virtually identical couplings to a resonant cavity mode.
The investigated situation is sketched in \figref{setup} a,
depicting an optical analog where three two-state atoms are
deterministically positioned at electric field antinodes of a cavity
mode where the coupling is maximum.
In our circuit QED \cite{Wallraff2004b,Schoelkopf2008} realization
of this configuration (\figref{setup} b), three transmon-type
\cite{Koch2007} superconducting qubits are embedded in a microwave
resonator which contains a quantized radiation field.
The cavity is realized as a coplanar waveguide
resonator with a first harmonic full wavelength resonance frequency
of $\omega_\textrm{r}/2\pi=6.729$~\textrm{GHz} and a photon decay
rate of $\kappa/2\pi=6.8$~\textrm{MHz}. The qubits are positioned at
the antinodes of the first harmonic standing wave electric field.
The transition frequency between ground $|g\rangle$ and first
excited state $|e\rangle$ of qubit $j$, approximately given by
$\omega_{\textrm{j}} \approx \sqrt{8
E_{\textrm{C}_\textrm{j}}E_{\textrm{J}_\textrm{j}}(\Phi_\textrm{j})}/\hbar
- E_{\textrm{C}_\textrm{j}}/\hbar$, is controllable through the flux
dependent Josephson energy
$E_{\textrm{J}_\textrm{j}}(\Phi_\textrm{j})=E_{\textrm{J
max}_\textrm{j}}|\cos{(\pi \Phi_\textrm{j}/\Phi_\textrm{0})}|$
\cite{Koch2007}. Here $E_{\textrm{C}_\textrm{j}}$ is the single
electron charging energy, $E_{\textrm{J max}_\textrm{j}}$ the
maximum Josephson energy at flux $\Phi_\textrm{j}=0$ and
$\Phi_\textrm{0}$ the magnetic flux quantum. Independent flux
control of each qubit is achieved by applying magnetic fields with
three external miniature current biased coils (\figref{sample} a)
where we take into account all cross-couplings by inverting the full
coupling matrix. Optical images of the investigated sample are
depicted in \figref{sample} b and c. The resonator was fabricated
employing optical lithography and Aluminum evaporation techniques on
a Sapphire substrate. All qubits were fabricated with electron beam
lithography and standard Al/AlO$_\textrm{x}$/Al shadow evaporation
techniques.  Table \ref{qubitparams} states the individual qubit
parameters obtained from spectroscopic measurements.

\begin{table}[b]
\begin{tabular*}{0.95 \columnwidth}[t]%
        {@{\extracolsep{\fill}}cccc}
    \hline
    Qubit $j$ & $E_{\textrm{C}_j}/h\ \textrm{(MHz)}$ & $E_{{{\textrm{J}_\textrm{max}}}_j}/h\ \textrm{(GHz)}$ & $g_\textrm{j}/2\pi \textrm{(MHz)}$\\
    \hline
    A & 283 & 224 & 83.7  \\
    B & 287 & 226 & -85.7  \\
    C & 294 & 214 & 85.1  \\
    \hline
\end{tabular*}
\caption{Qubit and qubit-resonator coupling parameters. The
single electron charging energy $E_{\textrm{C}_\textrm{j}}$, the
maximum Josephson energy
$E_{{\textrm{J}_{\textrm{max}}}_\textrm{j}}$ extracted from
spectroscopic measurements and the coupling strengths $g_\textrm{j}$
obtained from resonator transmission measurements for qubits $A$,
$B$ and $C$.} \label{qubitparams}
\end{table}

The physics of our system is described by the Tavis-Cummings
Hamiltonian \cite{Tavis1968}
\begin{equation}\label{tch}
\hat{\mathcal{H}}_{\textrm{TC}}= \hbar\omega_\textrm{r}
\hat{a}^\dagger \hat{a} + \sum_{j}{\left(
\frac{\hbar}{2}\omega_\textrm{j}\hat{\sigma}^{\textrm{z}}_\textrm{j}
+\hbar g_\textrm{j}(\hat{a}^\dagger \hat{\sigma}^-_{\textrm{j}}+
\hat{\sigma}^+_\textrm{j}\hat{a} )\right)} \, ,
\end{equation}
where $g_\textrm{j}$ is the coupling strength between the field and
qubit $j$. $\hat{a}^{\dagger}$ and $\hat{a}$ are the creation and
annihilation operators of the
field, $\hat{\sigma}^+_{\textrm{j}}$ and $\hat{\sigma}^-_{\textrm{j}}$
are the corresponding operators acting on the qubit $j$, and
$\hat{\sigma}^{\textrm{z}}_\textrm{j}$ is a Pauli operator. The
ground state $|g,g,g\rangle\otimes|0\rangle$ of the
three-qubit/cavity system is prepared by cooling the microchip to a
temperature of 20 \textrm{mK} in a dilution refrigerator.

\begin{figure*}[t]
\includegraphics[width=0.99 \textwidth]{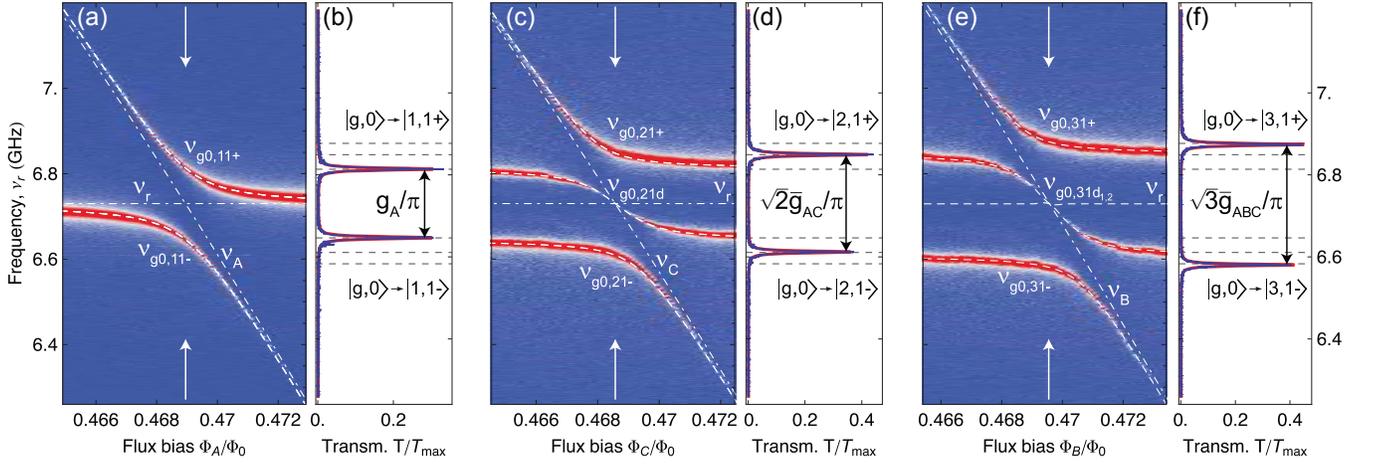}
\caption{Vacuum Rabi mode splitting with one, two and three
qubits. (a) Measured resonator transmission spectrum $T$ (blue, low
and red, high transmission) versus normalized external flux bias
$\Phi_\textrm{A}/\Phi_0$ of qubit $A$. Dash-dotted white lines
indicate bare resonator $\nu_\textrm{r}$ and qubit $\nu_\textrm{A}$
frequencies and dashed white lines are calculated transition
frequencies $\nu_{\textrm{g0,Nn}\pm}$ between $|g,0\rangle$ and
$|N,n\,\pm\rangle$. (b) Resonator transmission $T/T_{\textrm{max}}$
at degeneracy normalized to the maximum resonator transmission
$T_{\textrm{max}}$ measured at $\Phi_{\textrm{A,B,C}}=\Phi_0/2$ (not
shown), as indicated with arrows in (a). Red line is a fit to two
Lorentzians. (c) Resonator transmission spectrum $T/T_{\textrm{max}}$
versus external flux bias $\Phi_\textrm{C}/\Phi_0$ of qubit C with
qubit A degenerate with the resonator
($\nu_\textrm{A}=\nu_\textrm{r}$). (d) Transmission spectrum
$T/T_\textrm{max}$ at flux as indicated in (c). (e) Transmission
spectrum versus flux $\Phi_\textrm{B}/\Phi_0$ with both qubits A and
C at degeneracy ($\nu_\textrm{A}=\nu_\textrm{C}=\nu_\textrm{r}$).
The white dashed line at frequency
$\nu_{\textrm{g0,31d}_{\textrm{1,2}}}=\nu_\textrm{r}$ indicates the
dark state occurring at degeneracy. (f) Transmission spectrum
$T/T_\textrm{max}$ at flux as indicated in (e).} \label{data}
\end{figure*}

First we investigate the resonant coupling of the $|g\rangle$ to
$|e\rangle$ transition of qubit $A$ to the first harmonic mode of
the resonator. We measure the anti-crossing between qubit A
($\nu_\textrm{A}$) and the cavity ($\nu_\textrm{r}$) by populating
the resonator with much less than a photon on average. We record the
resulting transmission spectrum $T$ versus magnetic flux
$\Phi_\textrm{A}$ controlled detuning of qubit A (\figref{data} a).
Qubits B and C remain maximally detuned from the resonator at
$\Phi_{B}=\Phi_{C}=\Phi_0/2$ where they do not affect the
measurement. At finite detuning (left hand
side of \figref{data}a) we observe a shift of the resonator spectrum
which increases with decreasing detuning due to the dispersive
interaction with qubit $A$.

\begin{figure}[b]
\includegraphics[width=0.75 \columnwidth]{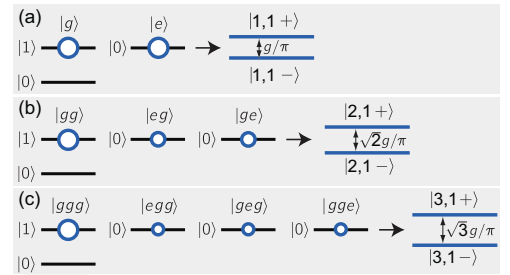}
\caption{Level diagram representing the total energy of
(a) one (b) two and (c) three
qubits resonantly coupled to a single photon. Bare energy
levels of the qubits $|g\rangle$, $|e\rangle$ and the cavity
$|0\rangle$, $|1\rangle$ are shown in black. The bright dressed
energy levels $|N,n\,\pm\rangle$, with $N$ the number of qubits,
$n$ the number of excitations and $\pm$ indicating the symmetry of the state, are illustrated in blue. The areas
of the circles indicate the relative population of the bare states in the
eigenstates $|N,n\,\pm\rangle$.} \label{levels}
\end{figure}

On resonance ($\omega_\textrm{j} = \omega_{\textrm{r}}$) and in the
presence of just one two level system ($N=1$), Eq.~\eqref{tch} reduces
to the Jaynes-Cummings Hamiltonian \cite{Jaynes1963}. The
eigenstates $|N, n\ \pm\rangle$ of this system in the presence of a single excitation
$n=1$ are the symmetric and anti-symmetric qubit-photon
superpositions $|1, 1\pm\rangle=\ 1/\sqrt{2}\
(|g,1\rangle \pm |e,0\rangle)$ (\figref{levels} a) where the
excitation is equally shared between qubit and photon. Accordingly,
we observe a clean vacuum Rabi mode splitting spectrum formed by the
states $|1,1 \pm\rangle$ (\figref{data} b). From analogous
measurements performed on qubits $B$ and $C$ (not shown) we obtain
the single qubit coupling constants $g_j$ listed in Tab.~\ref{qubitparams}.
The coupling strengths are virtually identical
with a scatter of only a few \textrm{MHz}. The strong coupling of an
individual photon and an individual two-level system has been
observed in a wealth of different realizations of cavity QED both
spectroscopically \cite{Wallraff2004b,Boca2004,Khitrova2006} and in
time-resolved experiments \cite{Brune1996,Hofheinz2008}. The regime
of multiple excitations $n$ which proves field quantization in these
systems has been reported both in the time resolved results cited
above and more recently also in spectroscopic measurements
\cite{Schuster2008,Fink2008,Bishop2008}.

In a next step, we maintain qubit A at degeneracy
($\nu_\textrm{A}=\nu_\textrm{r}$), where we observed the one-photon
one-qubit doublet (see left of \figref{data}c). Qubit B remains far
detuned ($\Phi_\textrm{B}=\Phi_0/2$) for the entire measurement.
Qubit C is then tuned through the already coupled states from lower
to higher values of flux $\Phi_\textrm{C}$. In this case, the
doublet states $|1,1 \pm\rangle$ of qubit A are found to be
dispersively shifted due to non-resonant interaction with qubit C
(\figref{data} c). When both qubits and the resonator are exactly in
resonance, the transmission spectrum T (\figref{data} d) shows only
two distinct maxima corresponding to the doublet $|2, 1\pm\rangle=\
1/\sqrt{2}\ |g,g\rangle\otimes|1\rangle \pm 1/2\
(|e,g\rangle+|g,e\rangle)\otimes|0\rangle$ with eigenenergies $\hbar
(\omega_\textrm{r} \pm G_\textrm{2})$. Here a single excitation is
shared between one photon, with probability $1/2$, and two qubits,
with probability $1/4$ each (\figref{levels} b). Both states have a
photonic component and can be excited from the ground state
$|g,g,g\rangle\otimes|0\rangle$ by irradiating the cavity with
light. These are thus referred to as bright states.
In general we expect $N+n=3$ eigenstates for two qubits and one
photon. The third state $|2,1 d\rangle=1/\sqrt{2}(|e,g\rangle -
|g,e\rangle)\otimes|0\rangle$ with energy $\hbar \omega_\textrm{r}$
at degeneracy has no matrix element with a cavity excitation and is
referred to as a dark state.
Accordingly we observe no visible population in the transmission
spectrum at frequency $\nu_\textrm{r}$ at degeneracy. In this regime
the two qubits behave like one effective spin with the predicted \cite{Lopez2007a}
coupling strength $G_\textrm{2}=\sqrt{2}\,\overline{g}_{\textrm{AC}}$
with $\overline{g}_{\textrm{AC}}=\sqrt{1/2(g_\textrm{A}^2 + g_\textrm{C}^2)}$,
which is indicated by dashed black lines in \figref{data} d. This
prediction is in very good agreement with our measurement.

%--------------------------------------------------------------------------------------------------------

Following the same procedure, we then flux tune qubit B through the
already resonantly coupled states of qubits A, C and the cavity
($\nu_\textrm{A}=\nu_\textrm{C}=\nu_\textrm{r}$), (\figref{data} e).
We observe the energies of three out of $N+n=4$ eigenstates, the fourth one being dark, for a range of flux values $\Phi_B$. Starting with
the dark state $|2,1 d\rangle$ at frequency $\nu_r$ and the doublet
$|2,1 \pm\rangle$ (left part of \figref{data} e), the presence of
qubit B dresses these states and shifts the doublet $|2,1
\pm\rangle$ down in frequency. Again one of these states turns dark
as it approaches degeneracy where it is entirely mixed with qubit B.
At degeneracy we identify two bright doublet states $|3,
1\pm\rangle=\ 1/\sqrt{2}\ |g,g,g\rangle \otimes |1\rangle\pm
1/\sqrt{6}\
(|e,g,g\rangle-|g,e,g\rangle+|g,g,e\rangle)\otimes|0\rangle$
(\figref{levels} c). The part of the states $|3,
1\pm\rangle$ carrying the atomic excitation is a so called
$W$-state, in which a single excitation is equally shared among all
$N$ qubits \cite{Dur2000}. Both $|3, 1\pm\rangle$ states are clearly
visible in the transmission spectrum shown in \figref{data} f.

In addition, there are two dark states $|3,1 d_1\rangle=\
1/\sqrt{2}(|e,g,g\rangle - |g,g,e\rangle) \otimes |0\rangle$ and
$|3,1 d_2\rangle=\ 1/\sqrt{2}(|g,e,g\rangle+|g,g,e\rangle) \otimes
|0\rangle$ which do not lead to resonances in the transmission
spectrum at degeneracy. In general all $N+n-2$ dark states are
degenerate at energy $\hbar \omega_\textrm{r}$.
The symmetries of the dressed three-qubit states are determined by
the signs of the coupling constants $g_\textrm{A}\approx
-g_\textrm{B}\approx g_\textrm{C}$.
While our measurement is not sensitive to the sign of coupling,
it is a simple consequence of the phase shift of the electric field mode by $\pi$
between the ends and the center of the resonator.
Again, the observed
transmission peak frequencies are in agreement with the calculated
splitting of the doublet
$G_\textrm{3}=\sqrt{3}\,\overline{g}_\textrm{ABC}$ (dashed black lines in
\figref{data} f). Also at finite detunings the measured energies of
all bright states are in excellent agreement with the predictions
based on the Tavis-Cummings model (dashed white lines in
\figref{data} a,c,e) using the measured qubit and resonator
parameters. We have also performed analogous measurements of all
twelve one, two and three qubit anti-crossings (nine are not shown)
and find equally good agreement.

\begin{figure}[t]
\includegraphics[width=0.50 \columnwidth]{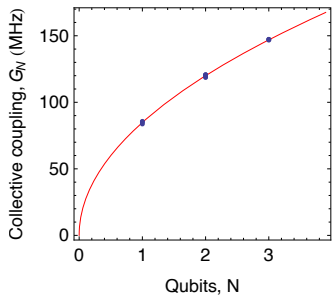}
\caption{Scaling of the collective dipole coupling strength.
Measured coupling constants (blue dots)
extracted from Fig.~\ref{data} and nine similar data sets and theoretical scaling (red line).} \label{data2}
\end{figure}

In \figref{data2} all twelve measured coupling strengths (blue dots) for one, two and
three qubits at degeneracy are plotted \textsl{vs.}~$N$.
Excellent agreement with the expected collective interaction
strength  $G_\textrm{N}=\sqrt{N}\,\overline{g}_{\textrm{ABC}}$ (red
line) is found without any fit parameters and
$\overline{g}_{\textrm{ABC}}=84.8$~\textrm{MHz}.

Our spectroscopic measurements clearly demonstrate the collective
interaction of a discrete number of quantum two-state systems
mediated by an individual photon. All results are in good agreement
with the predictions of the basic Tavis-Cummings model in the
absence of any number, position or coupling fluctuations. The presented
approach may enable novel investigations of super- and
sub-radiant states of artificial atoms.
Flux tuning on nanosecond timescales should furthermore allow the controlled generation
of Dicke states \cite{Stockton2004,Lopez2007b}
and fast entanglement generation via collective interactions
\cite{Tessier2003,Retzker2007}, not relying on individual qubit
operations. This could be used for quantum state engineering and an
implementation of Heisenberg limited spectroscopy
\cite{Leibfried2004} in the solid state.

%1444 words in main body

%-----------------------------------------------------
\begin{acknowledgments}
We thank T.~Esslinger and A.~Imamo\u{g}lu for discussions. This work
was supported by SNF grant no.~200021- 111899 and ETHZ. P.~J.~L.~was
supported by the EU with a MC-EIF. A.~B.~was supported by NSERC,
CIFAR and the Alfred P.~Sloan Foundation.
\end{acknowledgments}
%-------------------------------------------------------------------------

%-------------------------------------------------------------------------
\end{document}